\documentclass[a4paper,twocolumn,10pt]{esapub} 
\usepackage{times}
\usepackage{graphicx}
\usepackage{natbib}

\usepackage{aas_macros} 
\usepackage{multirow}
\usepackage{wasysym} 

\DeclareTextSymbol{\degrees}{OT1}{23}

\title{Constraints on the injection energy of positrons in the Galactic centre region}
\author[1]{P.~Sizun}
\author[1,2]{M.~Cass\'e}
\author[1]{S.~Schanne}
\author[1]{B.~Cordier}
\affil[1]{DAPNIA/Service d'Astrophysique, CEA~Saclay, F-91191 Gif-sur-Yvette Cedex, France}
\affil[2]{Institut d'Astrophysique de Paris, 98~bis Bd Arago, F-75014 Paris, France}

\begin{document}


\maketitle

\begin{abstract}
   Recent observations of the 511~keV positron-electron annihilation line in the Galactic centre region by the INTEGRAL/SPI spectrometer have stirred up new speculations about the origin of the large corresponding positron injection rate. Beyond astrophysical candidates, new mechanisms have been put forward. We focus on the annihilation of light dark matter particles and review the various gamma-ray radiation components related to such a source of mono-energetic positrons in addition to the 511 keV line itself. We study the influence of the degree of ionisation of the bulge on this radiation, and its possible effects on the observational constraints on the mass of the hypothetical light dark matter particle or the injection energy of a mono-energetic source of positrons in general.
\end{abstract}
\begin{figure}[!hbtp]
   \centering
   \fbox{\includegraphics[width=0.4\textwidth, height=0.4\textwidth]{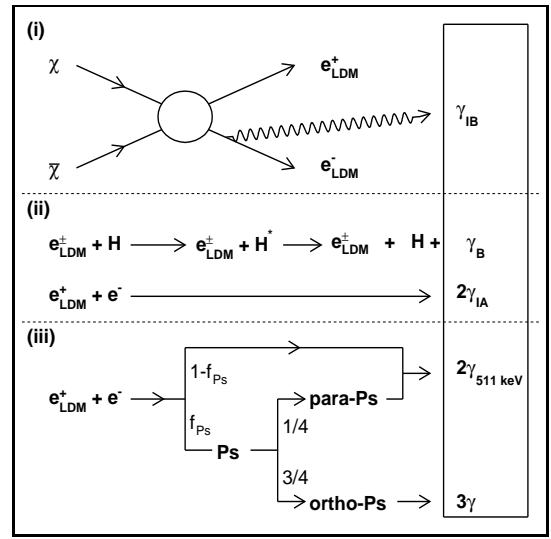}}
   \caption{Diagram illustrating the components of the gamma-ray radiation related to light dark matter. During the annihilation phase (i) of dark matter particles $\chi$, energetic positron-electron pairs are produced, and the Internal Bremsstrahlung radiation component is emitted. During the thermalization phase (ii), these positrons and electrons lose energy through their interactions with the ISM and Bremsstrahlung radiation is emitted; a fraction of the positrons annihilate in flight into two photons of variable energy. However, most of the positrons annihilate at rest (iii), either directly into a pair of 511~keV photons, or through the formation and decay of positronium.}
   \label{fig:diagramme}
\end{figure}
\section{Introduction}
Four years of observations with the INTEGRAL/SPI spectrometer 
 have permitted to precisely ascertain the flux --- $(1.07\pm{}0.03)\times{}10^{-3}~\mathrm{ph\,cm^{-2}\,s^{-1}}~$\cite{Jean2006}--- and the morphology ---$\sim{}8$\degrees{}~FWHM~\cite{Knodlseder2005}--- of the 511~keV gamma-ray line in the Galactic centre region discovered in 1978 by Leventhal et al.~\cite{Leventhal1978}, caused by the direct annihilation of positrons with electrons and the decay of para-positronium. Studies of the associated three gamma continuum from ortho-positronium decay revealed a similar morphology \cite{Weidenspointner2006} and lead to a positronium fraction of \mbox{$\mathrm{f_{Ps} = 0.967\pm{}0.022}$} \cite{Jean2006}.\\
These results motivated new reflections on the origin of the positrons responsible for this emission.  Because astrophysical candidates fell short explaining the $\sim{}1.4\times{}10^{43}~\mathrm{e^+/s}$ positron injection rate deduced from the 511~keV line flux, non nuclear candidates were considered, among which dark matter.\\
However, it was found that the gamma-ray radiation that would be produced if ``heavy'' dark matter particles such as the neutralino ($> 50~\mathrm{GeV/c^2}$) were the source of the positrons injected in the bulge, both immediately during the annihilation of dark matter particles and subsequently during the thermalization of the positrons, would be too high compared to CGRO/EGRET measurements. This is why an alternative possibility, \emph{i.e.} the self-annihilation of light dark matter (hereafter LDM) particles, with a mass $\mathrm{m_\chi}$ below $\sim{}200~\mathrm{MeV/c^2}$, was proposed \cite{Fayet2004,Boehm2004a}.\\
However, even such a scenario implies, besides the 511~keV line and the ortho-positronium continuum, additional gamma-ray radiation, as illustrated in Fig.~\ref{fig:diagramme}. It includes Internal Bremsstrahlung radiation during the annihilation of LDM particles \cite{Beacom2005a}, radiation from Bremsstrahlung interactions of the leptons with the inter-stellar medium and radiation from the in-flight annihilation of a small fraction of the positrons \cite{Beacom2006}.\\
Confronting this radiation with observations enables us to set constraints on the mass of the hypothetical LDM particle. We developed \cite[and references therein]{Sizun2006} a simple model of this radiation and derived upper masses based on CGRO/COMPTEL data. We summarize here the results, focusing on the influence on our constraints of the ionization fraction of the medium in which positrons thermalize.
\section{Model}
To model the gamma-ray radiation components associated to LDM annihilation, the first step consists in computing the steady-state spectrum of the positrons ---and electrons--- which are injected at a unique energy. It involves taking into account energy losses and in-flight annihilation through interactions with the inter-stellar medium.
\subsection{Positron and electron spectra}
\begin{figure}[!htbp]
   \centering
   \includegraphics[width=0.4\textwidth]{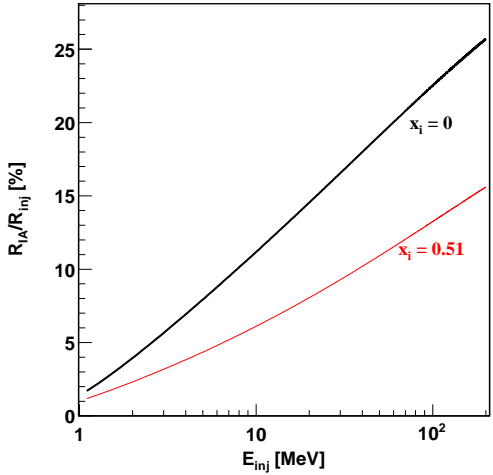}
   \caption{Fraction of positrons annihilating in flight as a function of the injection energy. The cases of a completely neutral propagation medium (black thick line) and of a medium with 51\% ionized phase (red thin line) are shown.}
   \label{fig:ria}
\end{figure}
\begin{figure}
   \centering
   \includegraphics[width=0.4\textwidth]{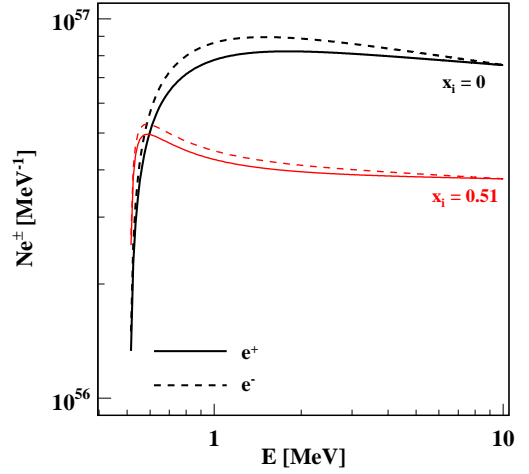}
   \caption{Steady state spectra of positrons (plain) and electrons (dashed) produced by the annihilation of light dark matter particles of mass 10~$\mathit{MeV/c^2}$ and propagating in a neutral medium (black thick lines) or in 51\% of ionized phase (red thin lines). The injection rate is tuned to ensure a 2$\gamma$ flux of $\mathit{1.07\times{}10^{-3}}$~$\mathit{ph/cm^2/s}$.}
   \label{fig:leptons}
\end{figure}
We start from a source of mono-energetic positrons and electrons, of total energy $E_{\mathrm{inj}} = \mathrm{m_\chi/c^2}$ injected in the Galactic bulge at a rate $R_{\mathrm{inj}}$:
\begin{equation}
	Q_{e^\pm\mathrm{inj}}(E) = R_{\mathrm{inj}} \delta(E-E_{\mathrm{inj}}). 
\end{equation}
We do not consider the morphology of the source, \emph{i.e.} the spatial distribution of dark matter particles, or the spatial propagation of these leptons within the bulge; we simply assume they are confined to the bulge. As they travel through the interstellar medium, they loose energy to the ambient medium and positrons can undergo in-flight annihilation.\\
The spectral distribution $N_{e^\pm}(E)$ of these leptons is obtained by solving the steady-state diffusion-loss equation
\begin{equation}
      \frac{d}{dE}\left(-N_{e^\pm}\frac{dE}{dt}\right) = -Q_{e^\pm}(E) \,,
 \label{eq:diffusion-loss}     
\end{equation}
where the key ingredients are the energy loss rate $\frac{dE}{dt}$ and the source term $Q_{e^\pm}$.\\
For electrons, the latter simply consists of the mono-energetic injection term, which yields the following distribution:
\begin{equation}
	\begin{array}{l@{\quad}l}
      	N_{e^-}(E) = \frac{R_{\mathrm{inj}}}{-dE/dt}	&	(mc^2<E<E_{\mathrm{inj}}).
	\end{array}      	
  \label{eq:electronspectrum}     
\end{equation}
For positrons, the source term also includes the in-flight annihilation (\emph{i.e.} before full thermalization) of a fraction (Fig.~\ref{fig:ria}) of the positrons injected:
\begin{equation}
     Q_{e^+}(E) = Q_{e^+\mathrm{inj}}(E) - n_e \sigma_{IA}(E)v(E)N_{e^+}(E) \,,
\label{eq:Q}
\end{equation}
where $\sigma_{IA}$ is the cross-section associated to in-flight annihilation. It results in a somewhat lower (Fig.\ref{fig:leptons}) steady state spectral distribution:
\begin{equation}
N_{e^+}(E) = \frac{R_{\mathrm{inj}}}{-dE/dt} e^{-\int_{E}^{E_{\mathrm{inj}}} n_e \frac{\sigma_{IA}(E')v(E')}{-dE/dt(E')}dE'}\,.
	\label{eq:positronspectrum}
\end{equation}  

\subsection{Radiation spectra}
\begin{figure}
   \centering
   \includegraphics[width=0.4\textwidth]{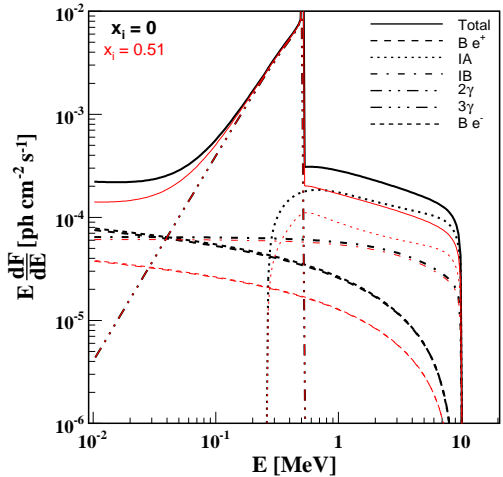}
   \caption{Radiation spectra related to 10~$\mathit{MeV/c^2}$ light dark matter particles, propagating in a neutral medium (black thick lines) or in 51\% of ionized phase (red thin lines). Components: positron annihilation spectra (in-flight:~IA; 2$\gamma$ at rest; 3$\gamma$ at rest), positron energy loss spectra (internal Bremsstrahlung: IB; external Bremsstrahlung: B $e^+$), electron Bremsstrahlung spectrum (B $e^-$) and total spectrum (plain).}
   \label{fig:components}
\end{figure}
The total rate of positron annihilation at rest depends on the initial positron injection rate $R_{\mathrm{inj}}$ and the in-flight annihilation fraction $R_{IA}/R_{\mathrm{inj}}$ given by our model. The respective fluxes of the 511~keV 2$\gamma$ line and the 3$\gamma$ ortho-positronium continuum are then determined using the positronium fraction $f_{Ps}=0.967\pm{}0.022$ \cite{Jean2006}. As we change the mass of the LDM particle and other parameters of the model such as the ionization fraction $x_i$ of the ISM, we tune the injection rate $R_{\mathrm{inj}}$ so that the 511~keV line flux remains equal to the observed value.\\
The other radiation components result from internal Bremsstrahlung during the LDM annihilation process, from Bremsstrahlung interactions of both positrons and electrons with the ISM during their thermalization and from the in-flight annihilation of some positrons into two photons. They are computed using the steady state spectra above and the appropriate cross-sections. More details are given in \cite{Sizun2006}.\\
Fig.~\ref{fig:components} displays the total and individual LDM radiation spectra for an arbitrary injection energy. Above the 511~keV line, in-flight annihilation dominates; Bremsstrahlung emission is negligible at high energies but, around a few tens of keV, it is of the same order as internal Bremsstrahlung
\subsection{Influence of ionization}
\begin{figure}
   \centering
   \includegraphics[width=0.4\textwidth]{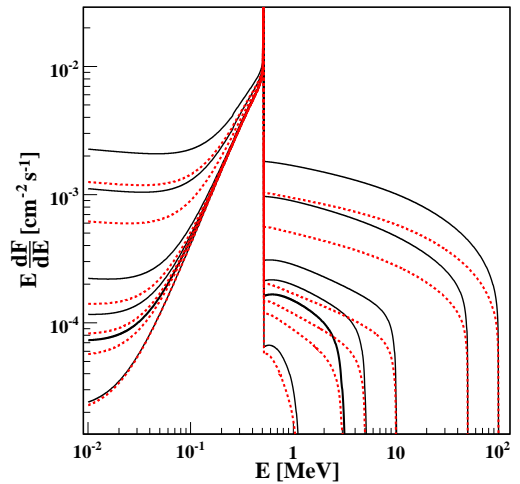}
      \caption{Total dark matter radiation spectrum for increasing injection energies $E_{\mathrm{inj}}$ (1, 3, 5, 10, 50 and 100~MeV), for a partially ionized (red dotted curves) or completely neutral (black solid curves) propagation medium.}
   \label{fig:spetot2}
\end{figure}

The ionization of the medium through which positrons propagate during their thermalization phase influences the in-flight annihilation and Bremsstrahlung radiation components. As $51^{+3}_{-2}\%$ of the rest annihilation takes place in the warm ionized phase of the ISM \cite{Jean2006}, we considered two cases for the ionization fraction $x_i = n_{HII}/n_H$ of the propagation region: a completely neutral medium and a medium made of 51\% of ionized phase.\\
Energy losses occur mainly through Coulomb and Bremsstrahlung interactions and Coulomb cross-sections are greater in the ionized phase than in a completely neutral environment. Consequently, the lepton spectra bump at lower energies (Fig.~\ref{fig:leptons}) and a smaller fraction of positrons annihilate in flight (Fig.~\ref{fig:ria}). In order to reach the same 511~keV line flux, a smaller injection rate is required which results in less Bremsstrahlung and in-flight annihilation radiation (Fig.~\ref{fig:components}-\ref{fig:spetot2}).
\section{Observational constraints}
\begin{figure}[!htbp]
		\includegraphics[width=0.4\textwidth]{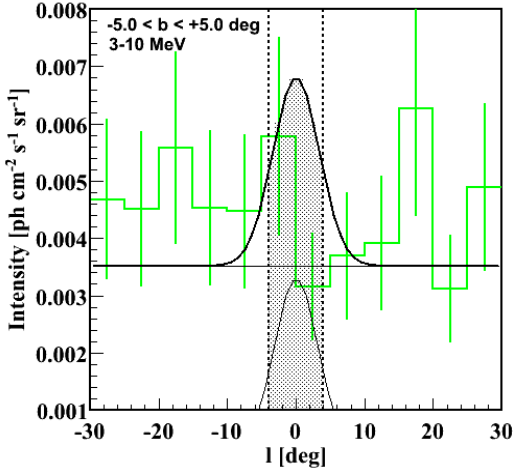} 
		\caption{Longitude profile of the 3-10 MeV COMPTEL skymap, in a latitude band of $\pm$5\degrees{} (histogram). The mean COMPTEL intensity in the inner disk region is used as a baseline and a possible excess in the Galactic bulge from LDM radiation is required to remain within data uncertainties.}
		\label{fig:profile}
\end{figure}
\begin{figure}[!htbp]
	\includegraphics[width=0.4\textwidth]{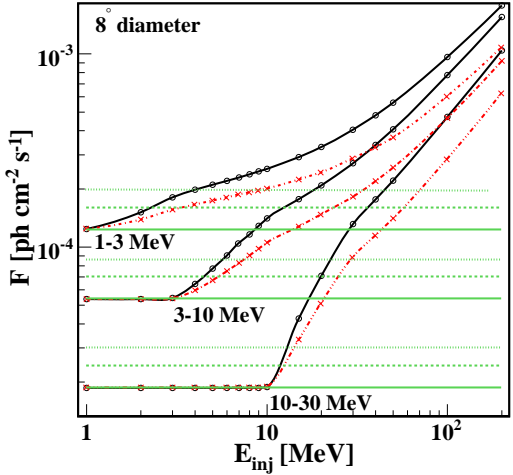}
	\caption{Total integrated flux in the 1-3, 3-10 and 10-30~MeV COMPTEL energy bands for an 8\degrees{} diameter region around the Galactic centre. Horizontal lines show the flux ---solid lines--- derived from the mean COMPTEL intensity in a wider fraction of the disk ($|l| < 30^\circ $, $|b|<5^\circ $) and its 1$\sigma$ ---dashed--- and 2$\sigma$ ---dotted--- upper values. The sum of the COMPTEL flux and of the LDM flux is shown as a function of the positron injection energy $E_{\mathrm{inj}}$, for a completely neutral propagation medium --- circled solid lines--- and for 51\% ionized medium --- crossed dot-dashed lines.}
	\label{fig:fluxtot}
\end{figure}

The CGRO/COMPTEL measurements of diffuse emission in the 1-30~MeV energy range \cite{Strong1999} provide the most stringent constraints, even though they shall be assigned systematic uncertainties of $\sim{}30\%$. Assuming the morphology of LDM gamma-ray continuum emission is similar to that of the 511~keV line, \emph{i.e.} is compatible with an 8\degrees{} FWHM gaussian, we consider a circular region of 5 or 8\degrees{} of diameter around the Galactic center and compute the flux expected from such a region (Fig.~\ref{fig:profile}) for various test injection energies. As no excess is observed in the COMPTEL skymaps in the Galactic bulge region with respect to the surrounding inner disk, we set our constraints by superposing our calculation to the mean COMPTEL flux and requiring that the total diffuse flux remains within data uncertainties (Fig.~\ref{fig:fluxtot}).\\
Resulting 2$\sigma$ upper limits on the positron injection energy are given in Tab.~\ref{tab:constraints}, both for the specific case of light dark matter and for the more general case of a source of mono-energetic positrons, \emph{i.e.} without the internal Bremsstrahlung component. As expected, the larger ratio of positrons surviving the thermalization phase within a partially ionized medium leads to slightly less constraining limits.

\begin{table}[!hbtp]
	\centering
	\caption{Constraints from COMPTEL data on the maximum positron injection energy $E_{\mathrm{inj}}$~[MeV]. 2$\sigma$ upper limits are given for an integration region of 5 or 8\degrees{} of diameter, for a neutral or 51\% ionized medium, with and without the Internal Bremsstahlung radiation component.}
	\label{tab:constraints}	
		\begin{tabular}[t]{|l|l|c||c|c|c|}
		\hline
		&	\multirow{2}{0.5cm}{\centering Size [\degrees{}]} & \multirow{2}{0.5cm}{\centering $x_i$ [\%]} & \multicolumn{3}{|c|}{Energy band} \\
		\cline{4-6}
		&		&	 & 1-3~MeV & 3-10~MeV & 10-30~MeV \\
		\hline
		\hline
			\multirow{4}{0.5cm}{\rotatebox{90}{with IB}} &	\multirow{2}{0.5cm}{\diameter~5} &
			  0    &	\textbf{3} & 5.5 & 12 \\
		\cline{3-6}		& &
			  51 & 7 & \textbf{6.5} & 13 \\
		\cline{2-6}
			&		\multirow{2}{0.5cm}{\diameter~8} &
			  0    & \textbf{4} & 6 & 12\\
		\cline{3-6}		& &			
				51 &	9 & 	\textbf{7.5} & 13.5 \\	
		\hline
		\hline
		\multirow{4}{0.5cm}{\rotatebox{90}{no IB}} &	\multirow{2}{0.5cm}{\diameter~5} &
			  0    &	\textbf{4} & 5.5 & 12.5 \\
		\cline{3-6}		& &
			  51   & 14 & \textbf{8.5} & 15 \\
		\cline{2-6}
			&		\multirow{2}{0.5cm}{\diameter~8} &
			  0    & \textbf{5.5} & 6.5 & 13.5\\
		\cline{3-6}		& &			
				51 &	15.5 & 	\textbf{10} & 16 \\	
		\hline
		\end{tabular}
\end{table}

\section{Conclusion}
By combining in-flight annihilation and internal Bremsstrahlung radiation, we conclude on a maximum mass of 3-4~$\mathrm{MeV/c^2}$ for the hypothetical LDM particle propagating within a neutral medium, in agreement with \cite{Beacom2006}.
Taking into account a possible ionization of the propagation medium enables us to release this constraint by a factor two. A scenario without internal Bremsstrahlung could even allow a source of 10~MeV mono-energetic positrons. A refined appraisal of the flux, morphology and spectral shape of diffuse emission in the 20~keV-20~MeV range, with an improved resolution and removal of point sources will be useful to fine-tune these upper limits. However, the mass range still allowed for the hypothetical light dark matter particle remains narrow.

\bibliographystyle{unsrt}
\begin{flushleft}
\bibliography{sizun2006iswm_v2}
\end{flushleft}

\end{document}